# The concentrated toroidal wave

Kevin J. Parker[1,2,3], Miguel A. Alonso[4,5]

[1]Department of Electrical and Computer Engineering, University of Rochester, 500 Computer Studies Building, P.O. Box 270231, Rochester, NY 14627, United States

[2]Department of Biomedical Engineering, University of Rochester, NY, 204 Robert B. Goergen Hall, P.O. Box 270168, Rochester, NY 14627, United States

[3]Department of Imaging Sciences, University of Rochester Medical Center, 601 Elmwood Ave, P.O. Box 648, Rochester, NY 14642, United States

[4]Aix Marseille Univ, CNRS, Centrale Med, Institut Fresnel, UMR 7249, Marseille Cedex 20, 13397, France

[5]Institute of Optics, University of Rochester, Rochester, NY 14627, United States

Email:  kevin.parker@rochester.edu; miguel.alonso@rochester.edu

**Abstract**

The classical solution to the Helmholtz wave equation in spherical coordinates is well known and has found many important applications in wave propagation, scattering, and imaging in optics and acoustics. The separable solution is comprised of spherical Bessel functions in the radial direction and spherical harmonics in the angular directions. The nature of the spherical Bessel functions includes a long asymptotic oscillatory tail at large radii, not conducive to applications where a tight concentration of wave amplitude around a ring is desired, for example in toroidal configurations. However, we have found that certain practical bandpass spectral shapes, centered around a peak frequency, can create a superposition of spherical Bessel functions that effectively concentrate the wave amplitude around a defined ring at the time instant of coherent addition, avoiding the long tail asymptotic oscillations of the single frequency solution. Theoretical solutions are shown for different bandpass spectra applied to the spherical Bessel functions, along with numerical solutions of transient wave propagation using practical hemispherical source shapes. These findings introduce a framework by which ring or toroidal concentrated waves can be produced with a simple bandpass superposition applied to hemispherical source shapes and with reference to the classical solutions in spherical coordinates.

# 1. Introduction

In many applications of optics and acoustics, there are advantages to using a tight focus, narrow beamwidth, or a concentrated area of maximum intensity. One particular form of concentration is related to the toroidal shape: a ring shape around a central coordinate. Solutions to acoustic and electromagnetic fields in toroidal coordinates have been derived despite the relative difficulty in working in a non-separable coordinate system [1, 2]. In some limiting cases, Weston's toroidal functions approach the spherical Bessel and harmonic functions which would be the separable solutions in 3D polar coordinates. In this paper, we return to the issue of toroidal patterns of waves, but with two major goals: first to seek a concentration of intensity around a narrow ring or toroid, without long sidelobe tails, and second goal to describe these within the classical framework of separable functions in spherical coordinates, with the advantages of separability in space and transform domains, and their relative simplicity.

# 2. Theory

In spherical coordinates, the free space Helmholtz equation is known to be separable [3-6] and can be written as:

$$P(r,\theta,\varphi) = \sum_{n=0}^{\infty} \sum_{m=-n}^{n} \left( a_{nm} j_n(kr) + b_{nm} y_n(kr) \right) Y_n^m(\theta,\varphi). \tag{1}$$

Here $j_l(kr)$ and $y_l(kr)$ are the spherical Bessel functions, $k$ is the wavenumber, $Y_l^m(\theta,\varphi)$ are the spherical harmonics [7], and $a_{nm}$ and $b_{nm}$ are the amplitudes assigned to each of the functions. In the summation limits, $n$ represents the integer orders of the spherical Bessel functions and $m$ represents the finite number of azimuthal angle orders. Note that this form provides general solutions, and requires boundary conditions to be specified to be used in any specific case [8, 9]. Here we use conventional physics notation for the spherical coordinates, as shown in **Figure 1**.

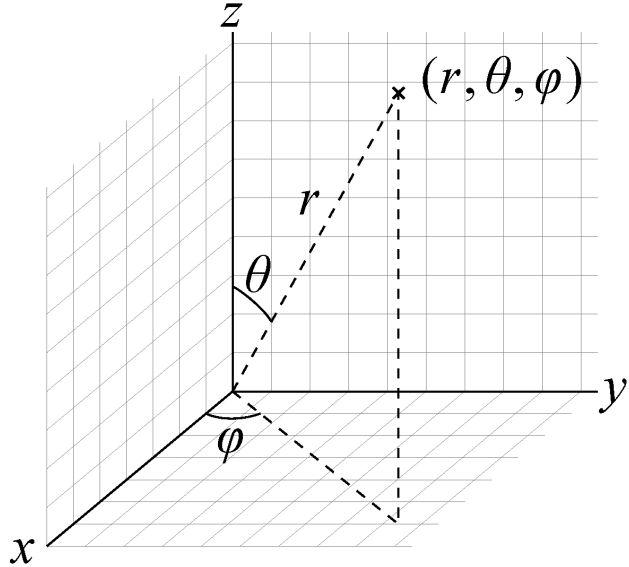

**Figure 1** Spherical coordinate system used with the spherical harmonic functions. We refer to the $x - y$ plane, where $z = 0$, as the equatorial plane. Figure reused with public domain permission [10].

Now to simplify these, let us restrict our examinations to the solutions that are finite at $r = 0$, and are comprised of a single integer order in $j$ and $Y$:

$$P_n(r,\theta,\varphi) = a_n j_n(kr) Y_n^n(\theta,\varphi). \qquad (2)$$

We examine in particular the $Y_n^n$ which are of the form $e^{(in\varphi)}(\sin(\theta))^n$ so are concentrated and reach their maximum at $\theta = \pi/2$, (the $x, y$ plane, or equator) and rotate around the $\varphi$ direction with $n$ cycles. The $j_4 Y_4^4$ is used in the following sections as a convenient example. Unfortunately, for applications that benefit from concentrated or focused beam patterns, the $j_4$ function of radius is not desirable, as its asymptotic decay is a "long tail" $1/r$ envelope [11]. However, we find that a broadband version of this can produce a field which is concentrated in time and space. Let us examine a superposition of $j_4$ patterns across a band of frequencies, with the spectrum $A(\omega)$, or alternatively $A(k)$ where the wavenumber $k = \omega/c$, and where $c$ is the wave speed. Initially, for the purpose of examining some simpler closed form solutions, let us assume that $A(k)$ has zero phase associated with a reference time of $t = 0$. Then, equation (2) is extended as an integration across a real spectrum:

$$P_4(r,\theta,\varphi,t=0) = \int A(k) j_4[k \cdot r] Y_4^4(\theta,\varphi) dk, \tag{3}$$

and because of the separable terms, the $Y_4^4$ term is independent of both $k$ and $r$, so we can focus on the first two terms in the integrand. Thus, we examine the following integration over a range of wavenumbers:

$$P_4(r,t=0) = \int A(k) j_4[k \cdot r] dk, \tag{4}$$

with particular attention to spectra $A(k)$ that result in compact and non-oscillatory functions of $r$.

As a first example let us specify the spectrum $A(k) = k^6 I_4(ak) K_4(bk)$, where $I$ and $K$ are modified Bessel functions [11], which by proper choice of $a$ and $b$ produces a limited bandwidth achievable in practical systems. Then, applying the integral theorem 16, section 6.578 of Gradshteyn and Ryzhik [12] we have:

$$\int_0^\infty k^{\upsilon+1} I_\upsilon(a \cdot k) K_\upsilon(b \cdot k) J_\upsilon(k \cdot r) dk = \frac{2^{3\upsilon}(abr)^\upsilon \Gamma\left(\upsilon + \frac{1}{2}\right)}{\sqrt{\pi}\left[(b^2 - a^2 + r^2)^2 + 4a^2 r^2\right]^{\upsilon + \frac{1}{2}}} \tag{5}$$

$$\left[\operatorname{Re} b > |\operatorname{Re} a| + |\operatorname{Im} r|; \quad \operatorname{Re} \upsilon > -\frac{1}{2}\right].$$

Next, using the identity [11] relating a spherical Bessel function of order $n$ to a regular Bessel function of order $n + 1/2$:

$$j_n(kr) = \sqrt{\frac{\pi}{2kr}} J_{n+\frac{1}{2}}(kr), \tag{6}$$

we apply equation (5) with specific order $\upsilon = 4.5$ as:

$$\int_0^\infty k^6 I_{4.5}(a \cdot k) K_{4.5}(b \cdot k) j_4(k \cdot r) dk = \frac{C \cdot r^4}{\left[\left(b^2 - a^2 + r^2\right)^2 + 4a^2 r^2\right]^5} \quad \text{for } a, r > 0, \quad (7)$$

where $C$ is a constant and the term $k^6 I_{4.5}(a \cdot k) K_{4.5}(b \cdot k)$ represents the broadband spectral shape of the waves under consideration, corresponding to $A(k)$ in equation (3). This product may not appear to be an obvious choice for a useful broad band spectrum, however it has a conventional band-limited shape with proper choice of $b \geq a$. The amplitude approaches zero at low frequency, peaks at mid-range, and then at large argument (high frequency and large $k$) its asymptotic expansion [11] is approximately $k^6 \exp[-(b-a)k]$, which asymptotically approaches zero. An example is given in **Figure 2**.

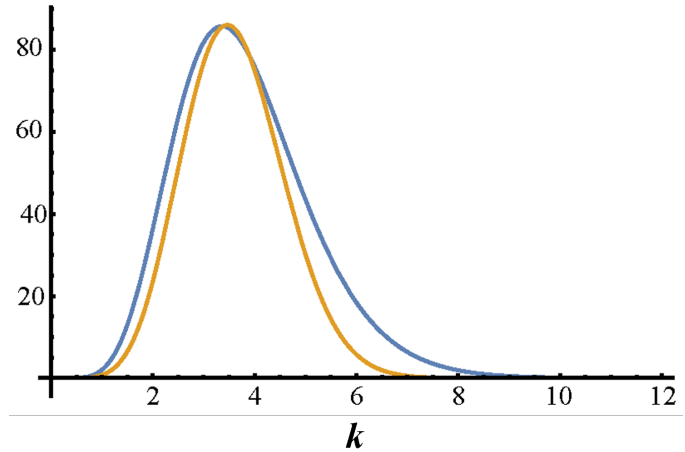

**Figure 2** Two spectra as a function of wavenumber $k$. Outer blue is $k^6 \text{BesselI}[4,5,k] \text{BesselK}[4,5,3k]$. Inner yellow is $k^6 \exp[-(1/4)k^2]$. Vertical axis is amplitude, arbitrary units. Both functions provide exact analytical solutions to a superposition of spherical Bessel functions of order $n$ resulting in toroidal-shaped waves concentrated around the origin of the spherical coordinate system.

The corresponding integral, the right side of equation (5) representing the superposition of $j_4$ functions across the specified spectrum, is also concentrated around a peak value, and has a large $r$ asymptotic approximation of $1/r^{(4\nu+(1/2))}$, a rapid decay with $r$ for reasonable orders of $\nu$, thus helping to produce a concentrated or spatially limited toroidal beam without extensive oscillations.

As a second example, we now examine the theorem 4 from section 6.631 of Gradshteyn and Ryzhik [12]:

$$\int_0^\infty k^{v+1} e^{-ak^2} J_v(rk) dk = \frac{r^v}{(2a)^{v+1}} \exp\left(-\frac{r^2}{4a}\right) \quad [\operatorname{Re} a > 0, \operatorname{Re} v > -1] \tag{8}$$

So as a specific example, let $v = 4.5$, and using equation (6) to convert to $j_4$, we have:

$$\int_0^\infty k^6 e^{-ak^2} j_4(rk) dk = \frac{e^{-\frac{r^2}{4a}} \sqrt{\pi} r^4}{64 a^{11/2}} \quad \text{for } a, r > 0,. \tag{9}$$

Thus, a bandlimited spectrum $A(k)$ of the form $k^6 e^{-ak^2}$ produces a function of $r$ that is of the form $r^4 e^{-r^2/4a}$, both compact by the properties of the Gaussian function. As a third example, we posit a case similar to the above but with an exponential bandlimited function $A(k)$ of the form $k^6 e^{-ak}$. With the theorems 1 and 2 from section 6.623 of Gradshteyn and Ryzhik [12] we have:

$$\int_0^\infty k^6 e^{-ak} j_4(rk) dk = \frac{\Gamma(6) 2^5 a r^4}{(a^2 + r^2)^6} \quad \text{for } a, r > 0. \tag{10}$$

This can also be extended to a more general time dependent case where the phase of the spectrum is assumed to be evolving over time as $e^{i\omega t}$, then from the same theorems 6.623 with an imaginary component on the exponential, explicitly written as $e^{i\omega t_0}$ and with normalized speed $c = 1$ the term is rewritten as a function of wavenumber $k$ in the integral, resulting in:

$$P_4(r,t) = \int_0^\infty k^6 \left(e^{ikt_0}\right)\left(e^{-ak}\right) j_4(rk) dk = \frac{\Gamma(6) 2^5 (a - it_0) r^4}{\left[(a - it_0)^2 + r^2\right]^6} \quad \text{for } a, t, r > 0. \tag{11}$$

Examples of this result, plotted as absolute value as a function of radius at three different time points are shown **in Figure 3**.

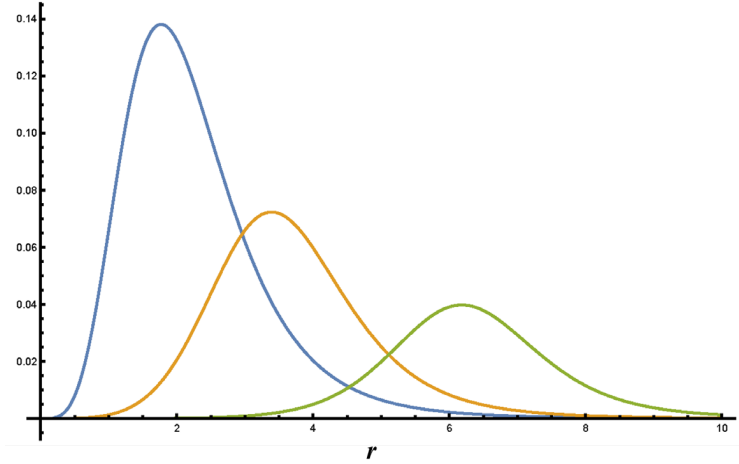

**Figure 3** The envelope of the complex waveform as a function of normalized radius (horizontal axis) at three different time points: reference time $t = 0$ (blue, largest magnitude), $t = 3$ (yellow), and $t = 6$ (green), in normalized time units and with the spectral parameter $a = 2.5$, showing outward propagation and drop in amplitude due to spherical spreading. Solutions are shown for the exponential broadband $j_4$ pulse of equation (11). Vertical axis is amplitude, arbitrary units.

## 3. Methods

All calculations were performed with Mathematica (version 13, Wolfram Research, Champaign, IL, USA) using their library of functions and graphics rendering. The numerical function "NItegrate" was used to apply Huygen's principle [13] of superposition for numerical field calculations of different examples and source configurations. Normalized units are used, specifically $c$, $\omega$, and $k$ are equal to unity.

## 4. Results

*4.1 The theoretical field solution for broadband $j_4 Y_4^4$*

First, we compare the free space solution of equation (2), single frequency vs. broadband, equation (7). **Figure 4** shows a 3D density plot of a half space using the transparent rendering style.

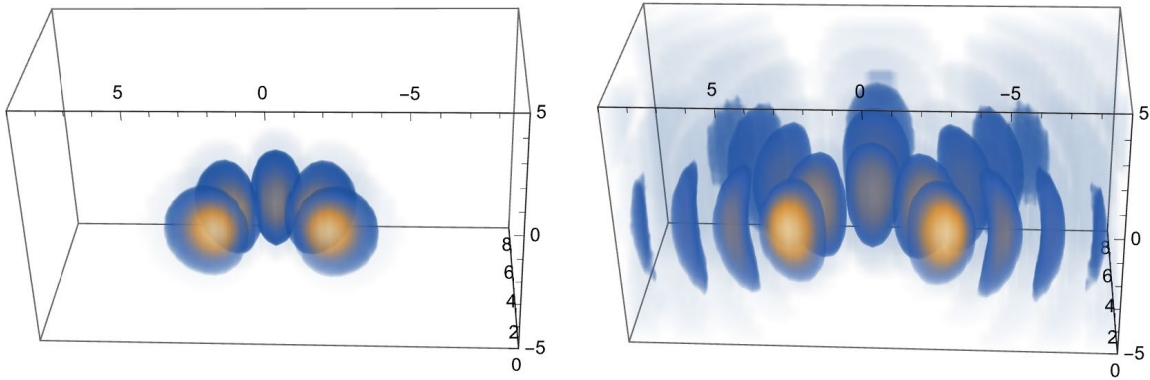

**Figure 4** 3D density plot of broadband (left) and narrowband (right) $j_4 Y_4^4$, 4-solutions in half-space, $z$-axis is vertical, origin is at the center of the facing surface. Wave intensity is represented as maximum mapped to bright yellow, minimum to dark blue then fading transparency towards zero. The broadband pattern is concentrated around the origin in the radial direction, unlike the narrowband solution. The broadband spectrum in this example is given by $k^6 \text{BesselI}[4.5, k] \text{BesselK}[4.5, 4.3k]$ and similar results are obtained for the $k^6$ Gaussian spectrum and $k^6$ exponential spectra. These provide a concentrated toroidal ring around the origin at reference time $t = 0$.

The wave distribution in **Figure 4** from the broadband will be a transient peak amplitude as the wave arrives from a hypothetical spherical source located outside of the figure. In laboratory settings, a large spherical source can be implemented for some specially bounded experiments in acoustics (sonochemistry, cavitation studies, radiation force, tweezers), optics, and electromagnetics. However, in many other applications we are restricted to a limited hemisphere or semi-hemisphere for positioning a source that can propagate waves into a target region. These conditions are examined in the next sections.

*4.2 Realization with a southern hemispherical source*

Now we examine the case of a hemispherical source (in the $-z$ or lower half plane) emitting according to a $Y_4^4$ pattern from a specific radius $r = 8$ normalized units. The solution in the *x-y* plane is of particular interest as it represents the peak intensity plane of the theoretical $j_4 Y_4^4$ pattern. This configuration is shown in **Figure 5** and the resulting broadband wave intensity at reference time $t = 0$ just above the equatorial plane are shown in **Figure 6**, demonstrating a toroidal pattern close to the full spherical solution of equations (3) and (5).

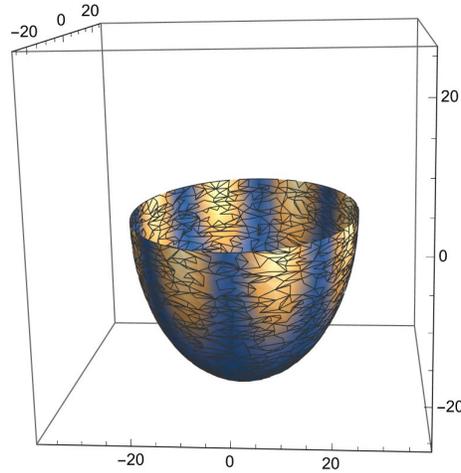

**Figure 5** Southern hemisphere source of $kr = 8$ normalized units with colors indicating strength of the $Y_4^4$ distribution. Source intensity is represented as maximum mapped to bright yellow, minimum to dark blue.

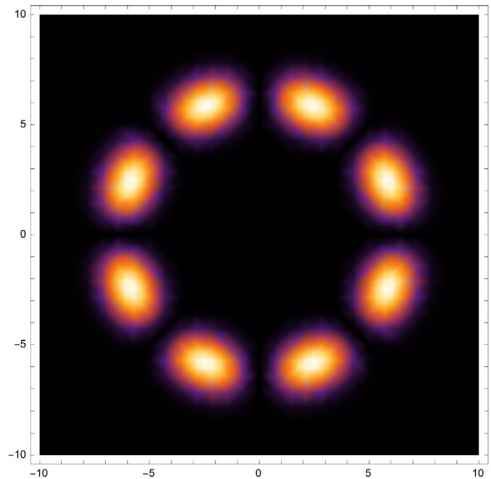

**Figure 6** Southern hemisphere broadband result, intensity distribution in the $(x, y, 0.1, t = 0)$ free space equatorial plane. Colors indicate wave intensity, represented as maximum mapped to white and bright yellow, minimum to dark blue then black near zero. This demonstrates the toroidal ring of the broadband $j_4 Y_4^4$ pattern at reference time $t = 0$.

*4.3 Realization with a western hemispherical source*

An alternative source configuration occupies the left-sided half space; this has a natural focus to the origin at the plane of maximum intensity of the $j_4 Y_4^4$ pattern. However the source configuration shown in **Figure 7** produces a broadband wave that at reference time $t = 0$ produces a toroidal pattern located at the $j_4 Y_4^4$ pattern indicated by **Figure 4(a)** and **Figure 6**, on the free space half plane (right side in **Figure 8**).

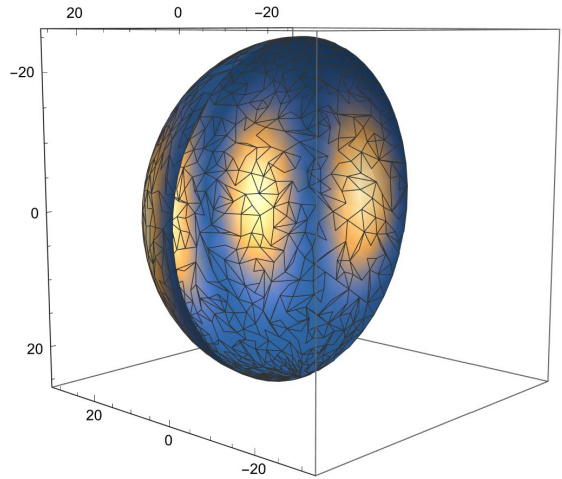

**Figure 7** Western hemisphere source of $kr = 8$ normalized units with colors indicating strength of the $Y_4^4$ distribution. Source intensity is represented as maximum mapped to bright yellow, minimum to dark blue. Note the edges are located at source minima of the $Y_4^4$ patterns.

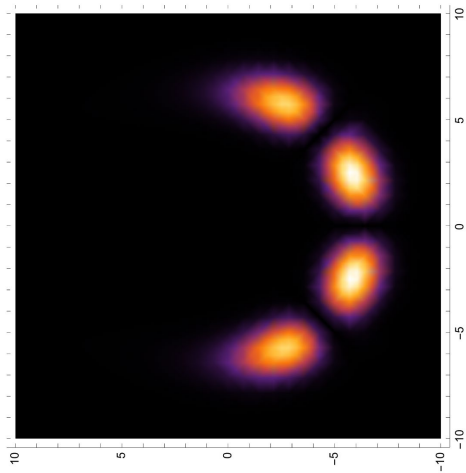

**Figure 8** Western hemisphere broadband result, intensity distribution in the $(x, y, 0.1)$ free space middle plane. Colors indicate wave intensity, represented as maximum mapped to white and bright yellow, minimum to dark blue then black near zero. This creates in the outer free space a half toroidal ring of the broadband $j_4 Y_4^4$ pattern at reference time $t = 0$.

*4.4 Realization with a lemon wedge-shaped source*

An interesting gambit concerns trimming back the area of the source transducer in the side or western hemispherical orientation to assess the practicality of a minimal footprint for the active source. Since the source function $Y_4^4$ approaches zero at the poles and at regular intervals around $\varphi$, we can isolate a wedge representing one cycle of modulation in $\varphi$ ($2\pi/n$) generally, and $n = 4$ in our examples) with amplitude at all edges approaching zero, generally an advantage in avoiding edge waves [14]. A source configuration covering one cycle of $Y_4^4$ in $\varphi$ is shown in **Figure 9**, and the time evolution of the broadband pulse produced by the spectrum indicated in equation (10) and equation (11) is shown in **Figure 10 (a-f)**. However, **Figure 10** is computed from numerical integration over the spatially limited source.

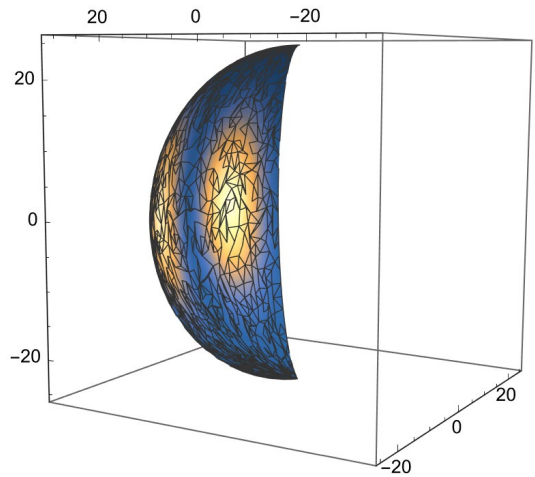

**Figure 9** "Lemon wedge" source of radius $kr = 8$ normalized units with colors indicating strength of the $Y_4^4$ distribution applied at the source surface. Source intensity is represented as maximum mapped to bright yellow, minimum to dark blue. Note the edges of this slice, which is a portion of the western hemisphere source, are located at source minima of the $Y_4^4$ patterns.

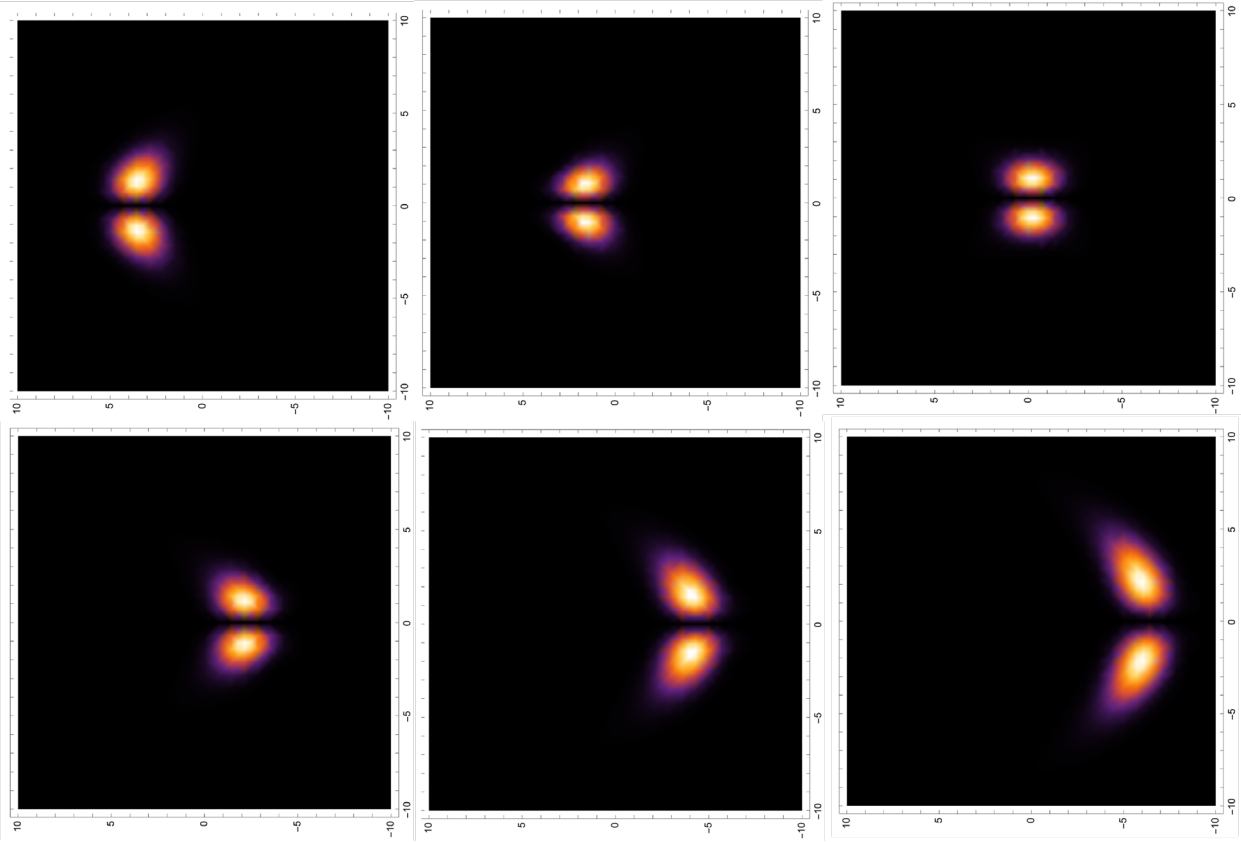

**Figure 10** Time evolution of broadband pulse from the "lemon wedge" source. Source located on the left and free space towards the right, numerical results are shown in the equatorial plane of $(x, y, 0, t = t_0)$. From top left, with respect to normalized time $t = 0$ and $c = 1$ (arbitrary units), we have $\Delta t = -10, -8, -6$. (top row) and then $\Delta t = -4, -2, \ 0$ (bottom row). Note that at the reference time 0, the wave is roughly in the position established by the full $j_4 Y_4^4$ pattern of **Figure 6**, where a full hemisphere would create 4 toroidal pairs, however the "lemon wedge" produces only one toroidal pair. Note also each figure is autonormalized to its own maximum.

## 5. Discussion

We have shown that it is possible to concentrate the generic $j_n Y_n^n$ waveform by using a specific bandpass spectrum; the superposition across some band of frequencies can create a toroidal shaped wave around the origin. There are some deeper mathematical issues behind this observation, linked to the spherical Hankel transform as described by Bracewell [15] and with three-dimensional properties described in detail by Baddour [16]. Using Baddour's notation we note that:

$$\hat{F}_n(\rho) = S_n\{f(r)\} = \int_0^\infty f(r) j_n(\rho r) r^2 dr. \tag{12}$$

"Note uses of the capital letter and of the hat to denote the spherical Hankel transform. $S_n$ is used to specifically denote the spherical Hankel transform of order $n$. The inverse transform is given by:"

$$f(r) = \frac{2}{\pi} \int_0^\infty \hat{F}_n(\rho) j_n(\rho r) \rho^2 d\rho. \tag{13}$$

Looking back at equations (9), we note that if we set $f(k) = k^n \exp[-k^2]$, then the spherical harmonic transform $S_n\{f(k)\}$ is $r^n \exp[(-r^2)/4]$. Thus, the function $k^n \exp[-k^2]$ is an eigenfunction of the spherical Hankel transform operator. The function, applied as a spectral shape, is also practical and advantageous. The practical nature stems from its band-limited shape as a function of frequency, and the advantages lie with the Gaussian term's compact concentration of energy. However, in order to invoke this eigenfunction relationship in the context of a band of spherical harmonics, the spectral shape needs to be of the form $k^{(n+2)} \exp[-k^2]$. Baddour's theorems also provide a framework for analytical treatments of truncation of the sources around the spherical geometry; these may ultimately yield analytical formulae for the hemispherical and semi-hemispherical examples in **Figures 7-9**; however this remains for future work.

It should be noted that generalization of our examples to other orders of Bessel functions and source shells of different radii are possible. We chose to work with examples from the $j_4 Y_4^4$ solution because of its simplicity, however other orders are possible according to the integral formulas in equations (5) – (11) related to Bessel function theorems. In general, as the order increases, the toroidal ring is located further out from the origin (following the general trend for the peak of the $j_n$ functions) and more oscillations are situated across the $\varphi$ circumference (following the trend for the spherical harmonics). Furthermore, in designing a source radius, we note that the angular spectrum approach of propagating waves from one plane to another has a direct analog in spherical coordinates [13] that propagates waves from one shell (of radius $r_1$) to another shell (of radius $r_2$), with preservation of form factors. Thus, the example of **Figure 8** may be thought of as concentrating the toroidal wave to the near surface of a semi-infinite medium on right side of the $z$-axis, (the positive $x$ space) with the source on the left as depicted in **Figure 7**. However, other larger shells can be configured and truncated with little effect since the source $Y_n^n$ pattern has negligible amplitude at the poles and periodically across $\varphi$. Alternatively, phased array techniques can be applied to refocus the location of the toroidal pattern deeper into the semi-infinite medium.

We note an interesting aspect of the sources examined in section 4, within a long reasonable range of radii, sources of different size (radius) will all produce the same resulting geometrical wave pattern so long as the appropriate $jY$ function and bandwidth are preserved on similar hemispherical sources of different radius. This is in marked contrast to the concept of f-number in aperture lens systems where increasing source radius creates a tighter focus. In our case the concentration is based on the separable $jY$ solution and not dependent on source radius, at least over some reasonable range from several lambda to very large multiples. Given these factors, there may be several strategies for creating concentrated toroidal patterns further out into free space, but these will require further research.

It should be noted that the examples of **Figures 5-10** are produced with sources that are quite small, the radius of each is $kr = 8$, so the hemisphere or semi-hemispheres are approximately 5/4 times the wavelength. Thus, small source footprints may be utilized in the production of these waveforms.

# 6. Conclusion

We have proposed several broadband versions of spherical harmonics that are capable of producing a field which is concentrated in time and space around a ring or toroidal pattern. The analytical solutions pertain to a fully spherical source, however we also show from numerical simulations that hemispherical and partial-hemispherical sources can provide similar results over limited angular extent. These fields and configurations may be useful in a variety of applications in optics and acoustics where a concentrated toroidal pattern described by the scalar wave equation is useful.

# CRedIT authorship contribution statement

**K. Parker:** Conceptualization, Methodology, Visualization, Writing – original draft, Writing – review and editing

**M. Alonso:** Conceptualization, Methodology, Writing – review and editing

# Declaration of competing interest

The authors declare that they have no known competing financial interests or personal relationships that could have appeared to influence the work reported in this paper.

## Data availability

Data available upon request.

## Acknowledgments

The authors thank their Departments for their supportive environments.

## ORCiDs

K J Parker https://orcid.org/0000-0002-6313-6605

M A Alonso https://orcid.org/0000-0001-7037-5383